\documentclass[12pt,letterpaper]{article}
\usepackage[dvips]{graphics}
\usepackage{graphics}
\usepackage[latin1]{inputenc}
\usepackage{amsmath}
\usepackage{amsfonts}
\usepackage{amscd,amssymb}

\usepackage{setspace}  
\usepackage[longnamesfirst,authoryear]{natbib}
\bibpunct{(}{)}{;}{a}{}{,}

\title{Does heterosexual transmission drive the HIV/AIDS epidemic in Sub-Saharan Africa (or elsewhere)? }
\author{Marc Artzrouni and Vivient Kamla\\
 Department of Mathematics\\
University of Pau\\
64013 Pau Cedex\\
France\\
Marc.Artzrouni@univ-pau.fr}

\begin{document}
\date{} 

\bibliographystyle{plainnat}

\maketitle

\begin{abstract} 
A two-sex  Basic Reproduction Number (BRN)  is used to investigate the conditions under which the Human Immunodeficiency Virus (HIV)  may spread through heterosexual  contacts in Sub-Saharan Africa. (The BRN is the expected number of new infections generated by one infected individual;  the disease spreads if the BRN is larger than 1).  A simple  analytical expression for the  BRN is derived  on the basis of recent data on survival rates, transmission probabilities, and levels of sexual activity.  Baseline results show that  in the population at large (characterized by equal numbers of men and women)  the BRN is larger than 1  if every year each person has 82 sexual  contacts with different partners.   the BRN is also larger than 1 for  commercial sex workers (CSWs)  and their clients (two populations of different sizes) if each CSW has about 256 clients per year and each client visits one CSW every two weeks. A sensitivity analysis explores the effect on the BRN of a doubling (or a halving) of the transmission probabilities. Implications and extensions are discussed. 
\bigskip

\end{abstract}

\begin{doublespace}
\newpage

 \section{Introduction}\label{intro}

There is a growing debate as to whether heterosexual contacts are the main mode of transmission of HIV-1 in Sub-Saharan Africa \citep{Gis03,Brew03, Fre06, Deu07}.   Some question the conventional wisdom  of a heterosexual epidemic on the basis that assumed  transmission probabilities per coital act are inflated \citep{Deu07}.  Others, on the contrary,  have developed complex mathematical models to show that alternatives  such as the use of unsafe medical injections were unlikely to be the main route of transmission because they rely on unfeasibly high iatrogenic transmission probabilities \citep{Fre06}.

In this paper we will show  that a complex mathematical model is not necessary in order  to assess  the feasibility of a heterosexual epidemic.  Indeed, the question can be studied by focusing on the basic reproduction number (BRN), which is  the expected number of \textit{secondary  infections} generated by one infected individual in a completely susceptible population (i.e. at  the beginning of an epidemic).    The disease will spread if and only if the $BRN$ is $ > 1$, i.e.  each infected  individual infects more than one other person.

The calculation of the BRN  hinges crucially on the evolution over the course of the infection of the  transmission probability  per coital act.   A  careful study of a population-based cohort of discordant couples (one person infected) in Rakai, Uganda, has shed light on this question  \citep{Gray01,Wa05}.   The first high-infectivity stage of the infection is characterized by an early peak in the viral load.  This pattern is paralleled by a rise in the transmission probability per coital act  that reaches a peak of about 0.008 before declining sharply one year into the infection.  During the long second (asymptomatic) stage,  the viral load is very low.  The probability of transmission remains also very low, at around 0.001 per coital act.  The third and last stage of the infection is characterized by a late  peak in the viral load (and in the probability of transmission).

The likelihood of a heterosexual epidemic depends on the transmission  probabilities but also on the number of partners.  For example,  with such relatively low  probabilities,  the disease may  not take hold  in a serially monogamous population, but could spread within high activity groups characterized by rapid changes in partnerships.    

For this reason it is sufficient to focus on the possible heterosexual spread between   high-activity groups. Indeed, if the disease can spread between such groups, it  will spill over to others  even in the transmission is inefficient from high to low activity sexual partners. The example that comes to mind is that of an epidemic that may spread efficiently between commercial sex workers (CSWs) and their  clients. The latter, in turn,  may infect, however inefficiently,  their  long-term female partners and thus spread the virus significantly among  low activity women.

We will first  derive an analytical  expression for the viral load on the basis of  recently available information obtained from the Rakai study  \citep{Gray01, Wa05}.   We will then use the log-log complementary model to obtain an expression for  the  probability of transmission per coital act.  This probability will be a function  (via the viral load)  of the infective age $ia$  (time since infection) and of the infective age at death $iad$  (time from infection to death).  We will call $ptr(ia,iad)$  this transmission  probability. 

Highly active groups (such as CSWs and their clients) will be characterized  by  annualized numbers of coital acts $NCA(ia,iad)$  (assumed to take place always with new partners).  This number depends on  the infective age and the  infective age at death because  the number of coital acts decreases as a person advances in the disease and approaches death \citep{Wa05}. 

The transmission rate for an individual who has been infected $ia$ years and will die at $iad$ years is now  $NCA(ia,iad)ptr(ia,iad)$.  If $s(x)$ is the density function of the infective age at death, the basic reproduction number $R_0$  is the expected value of the number of secondary infections generated by one individual during his/her infective life course (with maximum duration $\omega$): 

\begin{equation} \label{BRN}
R_0 \overset{def.}{ =} \int_{y=0}^{\omega} s(y)  \int_{x=0}^{y} NCA(x,y)ptr(x,y) dx dy.
\end{equation}

This is the BRN of a single sex model in which transmission is the same  between all individuals.  In particular the transmission rate and density of survival time are assumed to be the same for both sexes. However, survival differs slightly between men and women \citep{UN02}.   There can also be large differences in the number of coital acts  when dealing with groups of different sizes such as  CSWs and male clients.  Transmission rates may be different for the two sexes.  For example there is a growing consensus that male circumcision reduces significantly female-to-male transmission \citep{Nagel07}.      

For these reasons  we must consider two sex-specific basic reproduction numbers: the number $R_{fm}$ of secondary males infected by one infected woman ("the female to male BRN"), as well as $R_{mf}$, the male to female BRN.   The  product  $R_{fm}R_{mf}$  is  therefore  the  number of same-sex \textit{tertiary  infections} generated by one infected  person.   The expressions for $R_{mf}$ and  $R_{fm}$   will be those of Eq.  $\eqref{BRN}$ with the three  functions $s(y),  NCA(x,y)$ and $ptr(x,y)$   indexed by $m$ for $R_{mf}$ and by $f$ for $R_{fm}$.  

Although one could have defined the two-sex  BRN as the product  ${R_{fm}R_{mf}}$,  it  is generally defined as the  harmonic mean 
\begin{equation} \label{BRN1}
R_0 \overset{def.}{ =} \sqrt{R_{fm}R_{mf}}
\end{equation}
of the two sex-specific BRNs. This  definition   reflects the fact that transmission takes place over two generations \citep{Heest07}.  It is also consistent with the definition  of the BRN as the  dominant eigenvalue  of the next-generation matrix
$ \left(  \begin{matrix}     0 &    R_{mf} \\R_{fm}  & 0  \\   
\end{matrix}    \right) $. 

The threshold condition  for an epidemic flare-up is now $R_0>1$, i.e.  $R_{fm}R_{mf}>1$:  the number of same-sex tertiary infections generated by one infected individual must be larger than 1 for the epidemic to take hold. 

We will show that with a set of realistic  baseline parameter values, then in the population at large (characterized by equal numbers of men and women), the basic reproduction number $R_0$ is larger than 1 if  each year every person  has  82 sexual  contacts with different partners.  Within  the CSW-client populations (where men outnumber women)  the infection can spread  if each CSW has about 256 clients per year and each client visits one CSW every two weeks. 

The paper is organized as follows.  In Section 2 below we derive expressions for the viral load, the transmission probability per coital act, the annualized number of coital acts, and the density function of survival times. In Section 3 we give an expression for the two-sex basic reproduction number   and formulate the threshold condition in terms of the Index of Sexual Activity. Results are then illustrated with realistic  parameter values pertaining to Sub-Saharan Africa.  The sensitivity  of the results are discussed for different values of the probability of transmission function.   In Section 4 we discuss our findings, their implications and possible extensions. 

 \section{The four components of the basic reproduction number }\label{model}
The construction of the basic reproduction number is the same for both sexes.  For ease of exposition we will therefore drop the indexes   $f$ and $m$ from the functions (and parameters) used to define $R_{fm}$ and $R_{mf}$. 
\subsection{Viral load}\label{subvl}
In the absence of treatment,  the logarithm base 10 ($log_{10}$) of the viral load (measured in copies/mL)  follows  a well-established pattern as a function of the infective age $ia$  and of the infective age at death ($iad$)  \citep{Rapat05}.  During the  first year of infection the logarithm increases to approximately 5  and decreases rapidly thereafter  (first stage).  It then  remains around 3 during the  long asymptomatic second stage.  About  a year before death there is  a second peak in the  viral load.

We now describe a function  noted   $LVl(ia,iad)$, that captures this "twin peaks" pattern in the  $log_{10}$  of the viral load during the course of the infection.  The parameters that define the function are given in Table 1, together with baseline numerical values 
which reflect empirical results obtained from  the Rakai study \citep{Gray01, Wa05}. We take the same parameter values for both sexes.

\begin{singlespace}

 \bigskip
\begin{tabular}  {|l | l|}
  \hline
\multicolumn{2}{|c|}{Table 1: Parameters of $ LVl(ia,iad)$}\\ 
  \hline

   \it Parameter &  \it  Baseline value  \\
  \hline
 $ia_1$: Infect. age at first peak    &  0.4 year \\
  \hline
$M_1$: Value of  $LVl(ia_1,iad)$ at 1st peak & 5  \\
  \hline
$m$: Low value of $LVl(ia,iad)$  during second stage  &  3 \\
  \hline
 $\tau_1$: Time preceding death at  2nd peak    &  1 year \\
  \hline
$M_2$: Value of  $LVl(ia_1,iad)$ at 2nd peak & 4.8  \\
  \hline
 $\alpha_1, \alpha_2$: Parameters that determine the variance in the 1st peak    & 1.3; 0.2 \\
  \hline
 $\alpha_3$: Parameter that determines the variance in the 2nd peak    &  0.7 \\
  \hline
\end{tabular}  
 \bigskip
\end{singlespace}

We first define the  function
\begin{equation}
h_1(x,\alpha_1, M_1,ia_1) \overset{def.}{ =}   \dfrac{M_1 x^{\alpha_1-1}    exp \left[  \dfrac{x(1-\alpha_1)}{ia_1} \right]       }  {   ia_1^{\alpha_1-1}     \times exp (1-\alpha_1)}
\end{equation}
which reaches a maximum of $M_1$ for $x=ia_1$. We  then need the largest root $x^{\star}$ of the equation $h_1(x,\alpha_1, M_1,ia_1)=m$ in the unknown $x$;  $x^{\star}=1.647$ and will be used to obtain  the low value $m$ during the  long asymptomatic second stage. 

 We also need the function 
\begin{multline} 
h_2(ia,\alpha_2,x^{\star}) \overset{def.}{ =}  {x^{\star}[ 1+exp(-\alpha_2)]} \times \\  \left[              \left(  1+ exp\left[  \alpha_2 -   \dfrac{ia(1+exp(\alpha_2))}{x^{\star}}     \right]     \right)^{-1} - (1+exp(\alpha_2) )^{-1}    \right]         
 \end{multline} 
which will  be used as the argument $x$ in the function  $h_1$.  This will produce the first peak followed by the low value during the asymptomatic stage.  

The function 
\begin{equation}
h_3(x,y, \alpha_3,\tau_1) \overset{def.}{ =}     exp\left[   -\alpha_3(x-y+ \tau_1 )^2               \right]  
 \end{equation}
will be used to obtain the late-stage peak.  We combine these elements to finally  define 
\begin{multline} 
 LVl(ia,iad) \overset{def.}{ =}  h_1(h_2(ia,\alpha_2,x^{\star}), \alpha_1, M_1, ia_1)+  \\
  [M_2 - h_1(h_2(ia,\alpha_2,x^{\star}), \alpha_1, M_1, ia_1) ]h_3(ia,iad, \alpha_3, \tau_1).
\end{multline}

 The $log_{10}$ of the  viral load function  $LVl(ia,iad)$   corresponding to the parameter values in Table 1 is plotted in Figure 1  for an infective age at death ($iad$) of 7 years (together with the transmission probability  function derived below).  For a later infective age at death the function is similar with just a longer asymptomatic stage. 

\begin{figure}
\includegraphics{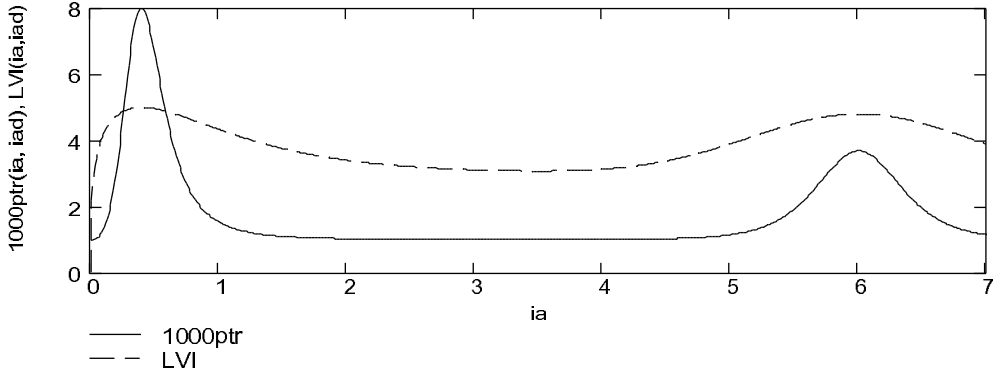}   
\caption{$log_{10}$ of viral load function  $LVl(ia,iad)$  and 1000 times probability of transmission per coital act function  ($1000ptr(ia,iad)$)  for a  person who dies seven years into the infection ($iad=7$).}
\end{figure}


\subsection{Transmission probability per coital act}\label{subprtr}
 \citet{Gray01} propose the log-log complementary model for the transmission probability per coital act  with (chronological) age and viral load as covariates (variables $age$ and $vl$).  Under this model this probability of transmission  is of the form 

\begin{equation}
ptr_0(age, vl)  \overset{def.}{ =}  1 - exp[-exp(\kappa_0 + \kappa_1vl +\kappa_2age)]
\end{equation}
with  parameters  $\kappa_m (m=0,1,2)$. 

The effect of age does not appear to be very strong \citep{Gray01, Wa05}  and would complicate the expression for the   basic reproduction number.  As  a simplification we therefore drop age as a covariate.  Bearing in mind that $10^{ LVl(ia,iad)}$ is the viral load,  we re-express an average (across ages) probability of transmission per coital act as the  function

\begin{equation}
ptr(ia,iad)  \overset{def.}{ =}  1 - exp[-exp(\kappa_0 + \kappa_110^{ LVl(ia,iad)})].
\end{equation}
We parameterize this function by specifying the values  $ptr_{hi}$ and  $ptr_{lo}$     of   $ptr(ia,iad)$   at the values $M_1$ and $m$  of $LVl(ia,iad)$  corresponding to the first 
peak in viral load and to the low plateau.    For given values of  $ptr_{hi}$ and  $ptr_{lo}$, the parameters $\kappa_0$ and $\kappa_1$ are then  the roots of the system
\begin{equation}
ptr_{hi} =   1 - exp[-exp(\kappa_0 + \kappa_110^{ M_1})],   \hspace{2mm}  ptr_{lo} =   1 - exp[-exp(\kappa_0 + \kappa_110^{ m})]
\end{equation}
from which 
\begin{equation}
\kappa_0=\dfrac{   ln    \left[     \dfrac {ln(1-ptr_{lo} )  }{ln( 1-ptr_{hi}) }  \right]  }     {10^{M_1-m}-1} +  ln \left[  ln(1-ptr_{lo})^{-1}     \right],  \hspace{2mm}  
\kappa_1=\dfrac{   ln    \left[     \dfrac {ln(1-ptr_{lo} )  }{ln( 1-ptr_{hi}) }  \right]  }     {10^m -10^{M_1}}.
\end{equation}
With the  numerical values $ptr_{lo}=0.001$ and $ptr_{hi}=0.008$, the resulting  function $ptr(ia,iad)$   (multiplied by 1000 in Figure 1)  provides  a good stylized approximation of  recent empirical estimates based on the Rakai study \citep{Wa05}.  We take the same parameter values for both sexes.

\subsection{Annualized number of coital acts with different partners}\label{subNCA}
We next construct a functional form for the annualized number of coital acts  $NCA(ia,iad)$. This function  will reflect a decreasing level of sexual activity as an infected person approaches death \citep{Wa05}.  The parameter $\Delta$ will be the value of $NCA$ at the time of infection ($NCA(0,iad)=\Delta$), i.e. the annual  number in the absence of HIV infection.  

The parameter  $ \phi$ will be the  fractional number of coital acts  remaining when an individual  reaches the infective age $iad-\tau_1$ at which the viral load reaches its second (pre-death)  peak ($NCA(iad-\tau_1,iad) =\Delta\phi$).    Finally $NCA$ will be 0 at the time of death ($NCA(iad,iad)=0 $). 
 \begin{figure}
 \includegraphics{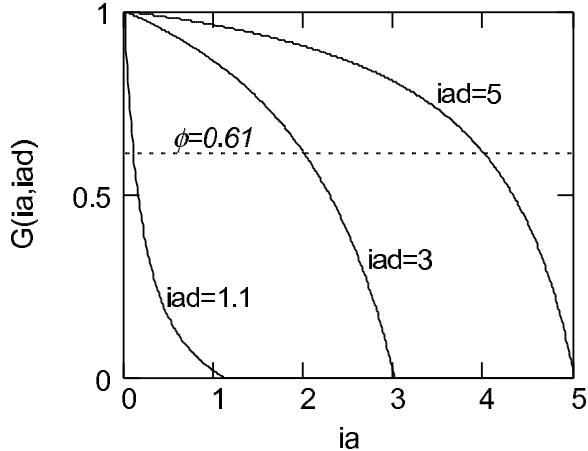} 
 \caption{Fractional number $G(ia,iad)$  of (annualized) coital acts remaining $ia$ years into the infection  for an  individual  who will die $iad$=1.1, 3 or 5 years into the infection (with $\tau_1=1$ and $\phi=0.61$, the function satisfies $G(0.1,1.1)=G(2,3)=G(4,5)=\phi=0.61)$.}
\label{labelfig2}
 \end{figure}

We now  define for $ia\le iad$ the function  $G(ia,iad)$ equal to the fractional number of (annualized) coital acts remaining for an  individual  infected $ia$  years ago and who will die $iad$ years  into the disease: 
\begin{equation}
 G(ia,iad)\overset{def.}{ =}           \left\{ \begin{array}{ll}
     \dfrac{ 1-ia/ iad }   {1+  \dfrac{ia(\tau_1-\phi.iad)}{iad.\phi(iad-\tau_1)}         }                      &       \mbox{ if }  iad > \tau_1                   ;\\
        0        &      \mbox{ if }  iad \le  \tau_1.\end{array} \right. 
\end{equation}
This function is equal to $1$ for $ia=0$, to $\phi$ for  $ia=iad-\tau_1$ and to 0 for $ia=iad$. 
A function $NCA(ia,iad)$  that has the required properties is obtained by multiplying $G(ia,iad)$ by $\Delta$: 
\begin{equation}
NCA(ia,iad)\overset{def.}{ =}   \Delta. G(ia,iad).
\end{equation}

 A baseline value of $\phi$ was taken equal to 0.61 on the basis of a mean reported number of coital acts per week of 10.2  at the beginning of the infection  and of a mean number  during a 6-15 month period prior to death of 6.2  (\citep{Wa05}, $10.2/6.2=0.61)$.   We take the same parameter values for both sexes. 

The function $G(ia,iad)$ is plotted  in Figure 2 for three different values of $iad$ and with $\tau_1=1$.  As $iad$ becomes closer to  $\tau_1$,  the function $G(ia,iad)$  of $ia$ approaches 0 more and more rapidly as $ia$ tends to $iad$. The fact that $G(ia,iad)$ is zero when the infective age at death $iad$ drops below  $\tau_1$  means that no sexual activity is assumed for a very short infection (e.g. an infection  that lasts less than one year  when $\tau_1=1$).  This drop to zero in sexual activity may not be entirely realistic, but is of little importance since there are extremely few, if any, infected individuals who will survive such a short period. 


\subsection{Density function of infective age at death $iad$}\label{dens}
Following the World Health Organization we assume a Weibull distribution for the infective age at death \citep{UN02}.   We parameterize this distribution with  its median  $me$ and shape parameter $\beta$. If we define $ \alpha \overset{def.}{ =}  me\left( ln(2)\right)^{-1/\beta}$       the density function $s(x)$ of $iad$  is then  

\begin{equation}
s(x)=\dfrac{x^{\beta-1}\beta}{\alpha^\beta}exp\left[    -\left(  \dfrac{x}{\alpha}  \right)^\beta      \right].
\end{equation}

We use $\beta= 2.5$  for the shape parameter for both sexes and a slightly shorter median for women ($me_f=8.6$ years) than for men ($me_m=9.4$ years) \citep{UN02}.

 \section{Results}\label{Res}
 \subsection{Threshold conditions on the basic reproduction number }\label{Th2}
In general  all  functions (and parameters)  are indexed  by $f$ and $m$.  The two  sex-specific  basic reproduction numbers are then 
\begin{equation}
R_{fm} \overset{def.}{ =} \Delta_f \int_{y=0}^{\omega}  s_f(y)  \int_{x=0}^{y} G_f(x,y)ptr_f(x,y) dx dy
\end{equation}
\begin{equation}
R_{mf} \overset{def.}{ =} \Delta_m\int_{y=0}^{\omega}  s_m(y)  \int_{x=0}^{y} G_m(x,y)ptr_m(x,y) dx dy. 
\end{equation}
 We next define the quantity 
\begin{equation}
I_0 \overset{def.}{ =} \left( \int_{y=0}^{\omega}  s_f(y)  \int_{x=0}^{y} G_f(x,y)ptr_f(x,y) dx dy \times  
\int_{y=0}^{\omega}  s_m(y)  \int_{x=0}^{y}  G_m(x,y)ptr_m(x,y) dx dy \right) ^{-1/2}.
\end{equation}
This quantity $I_0$ reflects at the individual level  the combined effects for  both sexes of variable infectivity, mortality, and  sexual activity over the course of the infection. 
We also define the  Index of Sexual Activity ($ISA$)  as the harmonic mean of the contact rates $ \Delta_m$ and   $\Delta_f$ (i.e.  the sex-specific annualized numbers  of coital acts at the beginning of the infection): 
\begin{equation}
ISA \overset{def.}{ =}    \sqrt{  \Delta_m \Delta_f }.
\end{equation}
 The Index of Sexual Activity measures the level of sexual activity between the two groups.

With these definitions,  the composite basic reproduction number 
\begin{equation}\label{BRN20}
R_0 \overset{def.}{ =} \sqrt{R_{fm}R_{mf}}
\end{equation}
will be larger than 1  if and only if 
\begin{equation} \label{Th1}
ISA > I_0.
\end{equation}\label{Th}
The (annualized) number of coital acts men have with women must be the same as the number of acts women have with men.  If $P_f$ and P$_m$ are the sizes of the corresponding  female and male populations, we must therefore have 
\begin{equation}
P_f \times  \Delta_f  = P_m \times  \Delta_m.
\end{equation}
The threshold condition $ \eqref{Th1}$  can then be paraphrased by saying that \textit{when both populations have the same size} then  $I_0$ is the minimum annualized number of coital acts each person must have with different partners in order for the disease to take hold  (since then $\Delta_f$ and  $\Delta_m$ are equal). 

 \begin{figure}
 \includegraphics{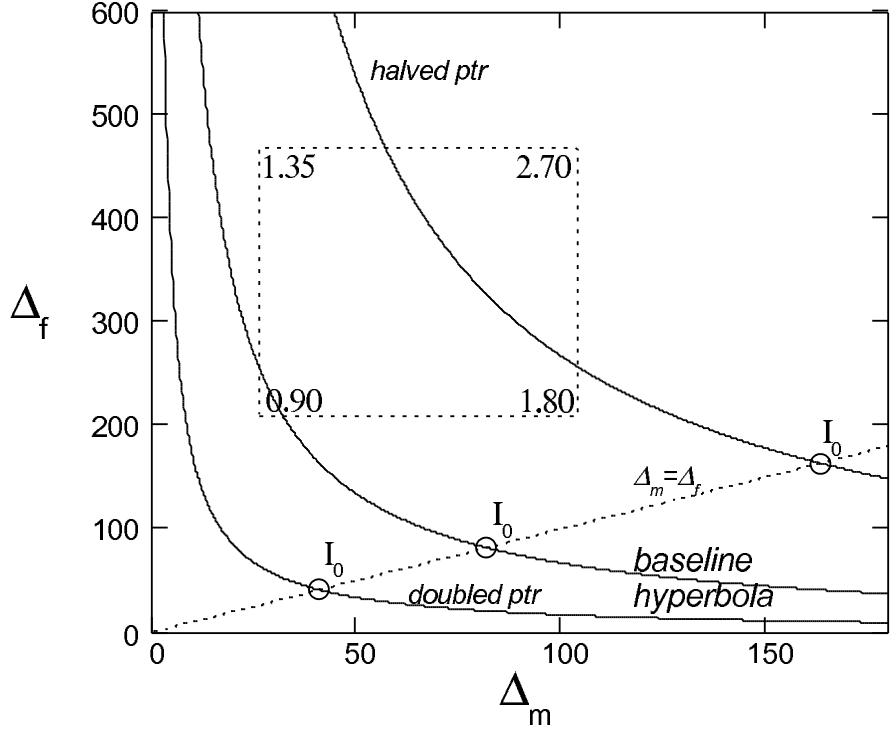} 
 \caption{Phase space of annualized  numbers  of coital acts by women ($\Delta_f$) and men ($\Delta_m$)  with locus  $\Delta_m\Delta_f=I_0^2 $  of values for which $R_0$ =1 (baseline hyperbola obtained with  parameter values of Section 2). The basic reproduction number $R_0$  is larger than 1 above the curve and vice-versa. The  black circle on the hyperbola  is the fixed point $I_0=81.60$, i.e.  the minimum  annual number of coital acts for the epidemic to spread when the male and female populations are of the same size. The "feasible rectangle" covers a range  of plausible values of  $\Delta_m$ and $\Delta_f$ for commercial sex workers and their clients (see text).   The  basic reproduction number at the four corners show that  $R_0$ is larger than one  in almost the entire feasible rectangle.   The two other hyperbolae correspond to  a halving and to a doubling of  the probability of transmission function $ptr$ for both sexes. The effect is linear on $I_0$  which is then 40.8 and 163.2. With a halving of the probabilities, $R_0$   is larger than one only for high levels of sexual activity (for example   $\Delta_m=100$  prostitute visits per year and $\Delta_f=400$ customers per year for each commercial sex worker).  With a doubling of the probabilities, $R_0$ is larger than one  in the entire feasible rectangle and well below.}
\label{labelfig3}
\end{figure}

\subsection{Numerical illustration (with baseline parameter values) }\label{Num}
The baseline parameter values and functions are those  given in Section 2.  They are the same for both sexes, except 
\begin {itemize}
\item for  the crucial sex-specific contact rates $\Delta_m$ and $\Delta_f$ between the two groups that will be used for the sensitivity analysis below.   
\item for  the slightly different median survival times  $me_f=8.6$ and  $me_m=9.4$. 
\end {itemize}

The quantity $I_0$ is independent of  $\Delta_m$ and $\Delta_f$ and its baseline value is 81.60.  In the ($\Delta_m,\Delta_f$) phase space  the corresponding baseline  hyperbola of equation  $\Delta_m\Delta_f=I_0^2$  is therefore the  locus of values  for which $R_0$ is equal to 1 (Figure 3).  

With male and female populations of the same size ($\Delta_m=\Delta_f$)  the fixed point  $I_0=81.60$ (black circle)   of the baseline hyperbola  tells us that each newly  infected person needs  about 82 coital acts per year with \textit{different}  partners in order for the disease to spread.  This is a small number compared to the documented  10.2 acts per week reported above.  The requirement that these acts take place with different partners, on the other hand, is in stark contrast with surveys that report  an average of about one partner per year in Sub-Saharan Africa \citep{Deu07}.  In short, the disease can spread between groups of men and women  of equal sizes for a reasonable annual number of coital acts, but with the requirement of a very high turnover of partners. 

Commercial sex workers and their clients are  groups of different sizes characterized  by a ($\Delta_m$, $\Delta_f$) point that lies   above the main diagonal  ($\Delta_m$=$\Delta_f$) of the phase-space diagram in Figure 3.  The number of acts per year varies considerably, however, with estimates for men in the range 0.5-2 prostitute visits per week, i.e.   $26 \le \Delta_m \le 104$ \citep{Nagel07}. A range for annual numbers of clients is based  on medians of 4 and 9 per week in rural and urban areas of Kenya \citep{Mee04}.  These medians  translate into the range $208 \le \Delta_f \le 468$  for annualized numbers of coital acts performed by each CSW with her clients.  The lower bound may reflect "casual" practices, while the upper one is probably conservative, with estimates of up  to 15 per day ($\Delta_f=5475$)  in Ghana \citep{Asa01}. 
The resulting "feasible rectangle" of ($\Delta_m,\Delta_f$)  values is depicted in Figure 3. The values of the basic reproduction number  $R_0$  (Eq. \eqref{BRN20}) at the four corners are given in the rectangle.  At the lower left corner each infected CSW will infect $R_{fm}=2.47$ clients, who will each in turn  infect $R_{mf}=0.33$ CSWs.  The resulting $R_0$ is equal to $\sqrt{R_{fm}R_{mf}}=0.90$. This shows that despite an efficient female-to-male transmission   the infection will not spread  if clients visit a CSW only every other week  and  CSWs service only 208 clients a year. If ($\Delta_m,\Delta_f$) moves up the left side of the rectangle then  $R_{mf}$ remains unchanged and the point  ($\Delta_m,\Delta_f$)     enters the   $R_0>1$ region for $\Delta_f=I_0^2/ \Delta_m=81.60^2/26=256.1$  (i.e. when $R_{fm}$ reaches $3.04$). For ($\Delta_m,\Delta_f$)  at  the upper right corner, $R_{fm}=5.55$ and $R_{mf}=1.32$: both BRNs are larger than 1 for an overall $R_0$ of $2.70$.

\subsection{Sensitivity analysis on transmission probabilities}\label{Sens}
There is a fair amount of uncertainly concerning the values of the peak and low   transmission probabilities  $ptr_{hi}$ and $ptr_{lo}$ whose baseline values were taken  as  0.008 and 0.001. For example the 0.008  figure was  an estimated probability at infective age 5 months, with a 95\% confidence interval of (0.004, 0.0015) \citep{Wa05}.   

In order to assess this sensitivity to the transmission probabilities we plotted  in  Figure 3   the two hyperbolae  corresponding to a halving (resp.  a doubling) for both sexes of both parameters $ptr_{hi}$ and $ptr_{lo}$.  This means a halving (resp.  a doubling) of the $ptr$ function and has a linear effect on $I_0$ which is  doubled to 163.2  (resp.  halved to 40.8).   With a halving of the $ptr$ function about two thirds of the feasible rectangle is in the $R_0<1$ region.  Fairly high contact rates $\Delta_m$ and $\Delta_f$  are needed in order for the disease to spread.  With a doubling of $ptr$ the rectangle is entirely in the $R_0<1$ region and  the spread will take place with much smaller values of $\Delta_m$ and $\Delta_f$. 

 \section{Discussion}\label{Dis}
Our goal was to use the  basic reproduction number $R_0$  to  investigate whether HIV-1 can spread in Sub-Saharan Africa (or elsewhere) primarily through heterosexual contacts.  \citet{Fra04} point out that published estimates of $R_0$ for generalized heterosexual HIV epidemics  are hard to come by.  This is no doubt because of complex transmission mechanisms and the heterogeneity of the populations involved.    

In this paper we have introduced a data-driven analytical  expression  for a two-sex basic reproduction number that captures  the nuances of variable infectivity and of changing levels of sexual activity  over the course of the infection.   



The results show that with our baseline parameter values the disease can spread between groups of men and women of equal sizes with a reasonable number of coital acts per year (82). However   these contacts must take place with different partners, which implies an unusually high level of sexual activity.  The infection   will spread between CSWs and their clients   for most plausible contact rates $\Delta_m$ and $\Delta_f$.  

By focusing on the BRN we were able to investigate conditions for a heterosexually driven HIV/AIDS epidemic on the basis of the survival, infectivity, and contact rates alone. We did  not have  to make  the many complex assumptions needed for a full-blown dynamic model of HIV transmission  (e.g. mixing patterns, partnership formation rules, etc.). Valuable insights have thus been  gained with a  minimum number of assumptions and parameters. 

Epidemiologists, public health officials and others can now make judgements concerning the potential spread of HIV-1 by  checking   the simple condition  $\Delta_m\Delta_f>I_0^2$  with local values  of  the contact rates  $\Delta_m$ and $\Delta_f$.  It is wise however to take into account the uncertainty on $I_0$ by considering its plausible  lower and higher values of 40.8 and 163.2.

At least the overall (average)  transmission probability obtained from  the Rakai study is  believed to be similar to that reported from "prospective studies of  European, north American and Thai heterosexual couples" \citep[p. 1152]{Gray01}.  The transmission function $ptr$  may therefore  be  applicable to populations outside of Uganda.  

Our results  assume no  intervention (condoms, circumcision, therapy, etc).  However the survival,  viral load or transmission  functions can be changed to assess the effect on the basic reproduction number of  longer survival and/or antiretroviral therapies.    More generally,  the theoretical results of Section 3.1 can be used for any sexually transmitted infection by using the appropriate function needed to calculate $R_{fm}$ and $R_{mf}$.

Finally we emphasize that our results are based on the assumption of instantaneous partner changes which  implies a high level of sexual activity.  Our results  are not applicable to a population   characterized by alternating periods in and out of a partnership \citep[as in][]{Mor00}.  Everything else being equal (survival, infectivity, etc.) expected numbers  of  secondary infections will be lower in such  partnerships.  This is  because during the whole duration of a (monogamous)  partnership an infected individual can transmit the virus to only one person.  If that happens, there will be  no more  transmission until the end of the current partnership, itself often  followed by a  period without sexual activity.  Calculating an expected number of secondary (and tertiary)  infections in this case is a difficult statistical problem currently under investigation.

\end{doublespace}

\end{document}